\documentclass[final]{IEEEcsmag}

\ifCLASSOPTIONcompsoc
  \usepackage[nocompress]{cite}
\else
  \usepackage{cite}
\fi
\ifCLASSINFOpdf
\else
\fi
\usepackage{comment}

\usepackage[dvipsnames]{xcolor}

\newcommand{\mytitle}{Quantum Computing and Visualization: A Disruptive Technological Change Ahead}

\newcommand{\fix}[1]{\textcolor{red}{\textbf{\textit{#1}}}}

\newcommand{\qx}{QC}

\newcommand{\qubit}{\emph{qubit}}

\newcommand{\qubits}{\emph{qubits}}
\newcommand{\Qubits}{\emph{Qubits}}

\newcommand{\phasedisk}{Phase Disk}
\newcommand{\blochsphere}{Bloch Sphere}
\newcommand{\blochspheres}{Bloch Spheres}
\newcommand{\qs}{Q-sphere}

\usepackage{physics}

\usepackage{caption}
\usepackage{subcaption}
\usepackage{hyperref}
\usepackage{amssymb}

\begin{document}


\title{\mytitle{}}

\author{E. Wes Bethel{\IEEEauthorrefmark{1},\IEEEauthorrefmark{2}}
Mercy G. Amankwah\IEEEauthorrefmark{3},
Jan Balewski\IEEEauthorrefmark{2}, Roel Van Beeumen\IEEEauthorrefmark{2}, 
Daan Camps\IEEEauthorrefmark{2}, 
Daniel Huang{\IEEEauthorrefmark{1},\IEEEauthorrefmark{2}},
Talita Perciano\IEEEauthorrefmark{2} \\
\IEEEauthorrefmark{1}San Francisco State University, 
\IEEEauthorrefmark{2}Lawrence Berkeley National Laboratory
\IEEEauthorrefmark{3}Case Western Reserve University}

\begin{abstract}
 
The focus of this Visualization Viewpoints article is to provide some background on Quantum Computing (\qx{}), to explore ideas related to how visualization helps in understanding \qx{}, and examine how \qx{} might be useful for visualization with the growth and maturation of both technologies in the future.
In a quickly evolving technology landscape, \qx{} is emerging as a promising pathway to overcome the growth limits in classical computing.
In some cases, \qx{} platforms offer the potential to vastly outperform the familiar classical computer by solving problems more quickly or that may be  intractable on any known classical platform. 
As further performance gains for classical computing platforms are limited by diminishing Moore's Law scaling, \qx{} platforms might be viewed as a potential successor to the current field of exascale-class platforms.
While present-day \qx{} hardware platforms are still limited in scale, the field of quantum computing is robust and rapidly advancing in terms of hardware capabilities, software environments for developing quantum algorithms, and educational programs for training the next generation of scientists and engineers.
After a brief introduction to \qx{} concepts, the focus of this article is to 
explore the interplay between the fields of visualization and \qx{}.
First, visualization has played a role in \qx{} by providing the means to show representations of the quantum state of single-\qubits{} in superposition states and multiple-\qubits{} in entangled states. 
Second, there are a number of ways in which the field of visual data exploration and analysis may potentially benefit from this disruptive new technology though there are challenges going forward. 
%

\end{abstract}

\maketitle


%
\IEEEpeerreviewmaketitle



\begin{figure*}[t!]
     \begin{subfigure}[b]{0.49\textwidth}
         \centering
         \raisebox{0.25in}{\includegraphics[width=\textwidth]{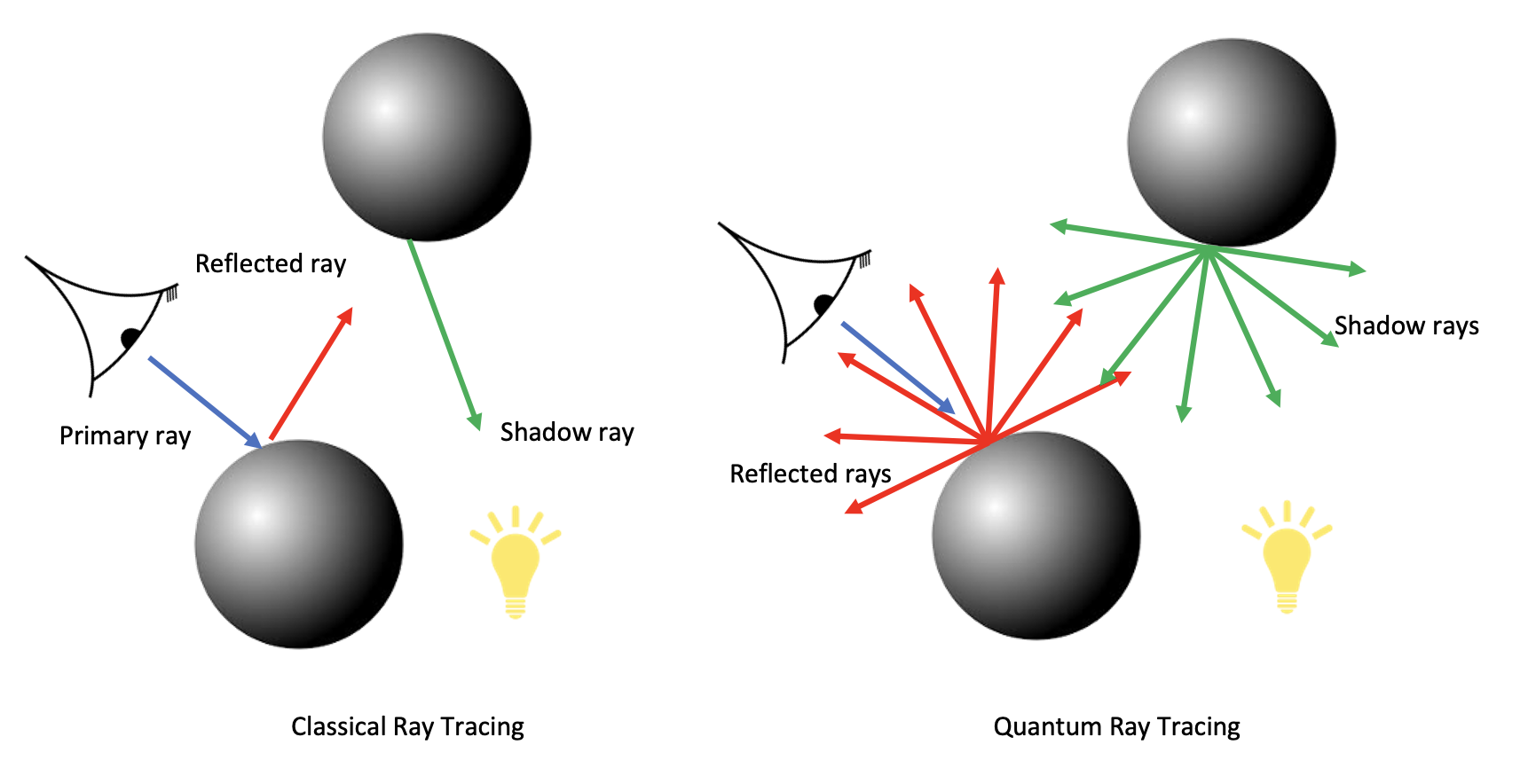}}
         \caption{In classical raytracing, a single ray is tested against objects in the scene one after another. In the quantum world, a single ray is tested against all objects in the scene at once. Image adapted from~\cite{QuantumRayTracing:2022}.}
         \label{fig:summary:classical_vs_quantum_ray_tracing}
     \end{subfigure}
     \hfill
     \begin{subfigure}[b]{0.45\textwidth}
         \centering
         \includegraphics[width=\textwidth]
         {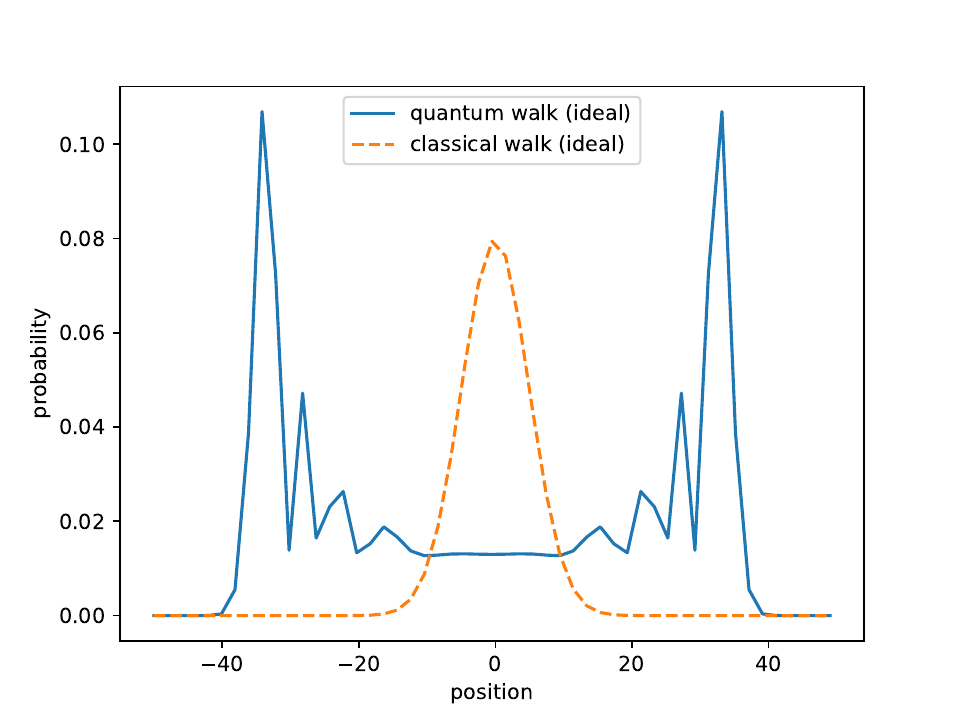}
         \caption{Classical random walks and quantum walks produce  different probability distributions resulting in vastly different sampling quality characteristics. }
        \label{fig:summary:classical_vs_quantum_walk}
     \end{subfigure}
     \label{fig:classical_vs_quantum_2}
     \caption{Quantum algorithms exhibit fundamentally different behavior from classical ones. }
\end{figure*}

\section{Introduction}

%
%
%
%

Visualization as we commonly know it provides a means for "seeing the unseeable", for transforming abstract information into readily comprehensible images~\cite{McCormick:1987}. The field of visualization has evolved within the rubric of classical computing constructs, where information is represented as binary values, zeros and ones, and manipulated using boolean and arithmetic operations.
Prior work in classical visualization is diverse and ranges from reformulating serial algorithms to perform better on parallel hardware, to incorporating the principles of optics and radiation transport modeling to create photorealistic renderings of models and data. 

The emerging field of \emph{quantum computing} (\qx{}) is shaping up to be significantly more disruptive than previous changes we've been through with classical computing technology. These technologies include the physical means for storing information, the mathematical basis for conceptually manipulating that information, and the semantic and syntactic way of expressing algorithms for implementing those manipulation concepts. 
Because \qx{} is so different than classical computing in so many respects,
we expect the transition into \qx{} to be even more disruptive than any of the technological transitions we have seen so far. 
 This thought applies broadly to all of computing as well as specifically to the field of visualization.
As the field of visualization continues to evolve, we may wonder  how it will be impacted by \qx{} and how it will impact \qx{}.

While the transition to \qx{} will likely be bumpy, 
there are several aspects of \qx{} that offer potential gains for graphics and visualization in terms of scalability of methods and quality of results. 
Fig.~\ref{fig:summary:classical_vs_quantum_ray_tracing} illustrates a key difference between classical and quantum raytracing. In the classical implementation, a ray is tested against objects in the scene one after another.
In the quantum implementation, a ray is tested against all objects in the scene with a single quantum operation applied at once to the $2^n$ quantum states created from $n$ entangled \qubits{}. 
Fig.~\ref{fig:summary:classical_vs_quantum_walk} shows how a classical random walk converges to a probability distribution that looks like a Gaussian. 
On the other hand, the quantum walk produces a distribution that more thoroughly covers the mathematical space~\cite{Kempe:2003}  resulting in a potentially higher quality result for algorithms that use randomized methods based on walks. 


The objective for this article is to explore how visualization has helped understanding \qx{} concepts (\S\ref{sec:vis_for_quantum}) as well as to consider ways that visualization may benefit from \qx{} technology (\S\ref{sec:quantum_for_vis}).
Because \qx{} is so much different than classical computing, we first provide a brief introduction of key ideas and terminology (\S\ref{sec:qc_background}).

\begin{figure*}
\centering
   \begin{subfigure}[t]{0.45\textwidth}
         \centering
         \raisebox{0.20in}{\includegraphics[width=\textwidth]{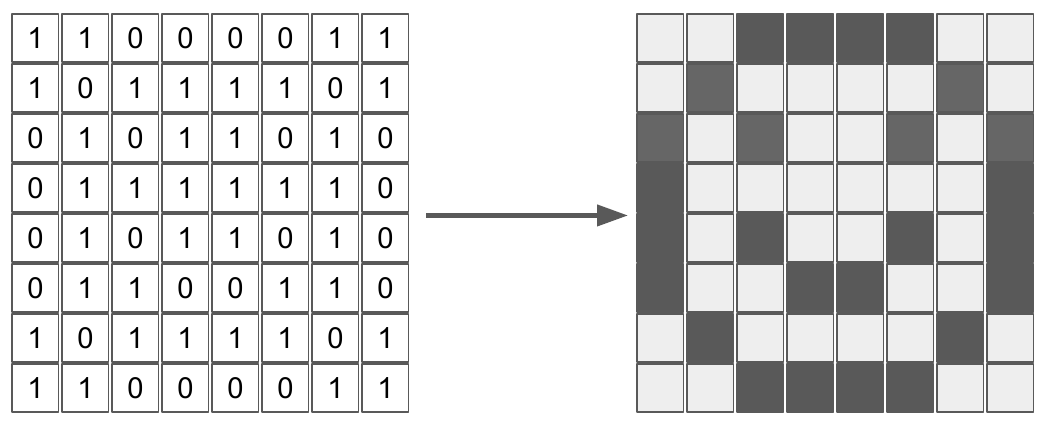}}
         \caption{Visualizing the state of 64 classical bits}
         \label{fig:summary:vis_classical_bits}
     \end{subfigure}
     \hfill
    \begin{subfigure}[t]{0.45\textwidth}
         \centering
         \includegraphics[width=\textwidth]{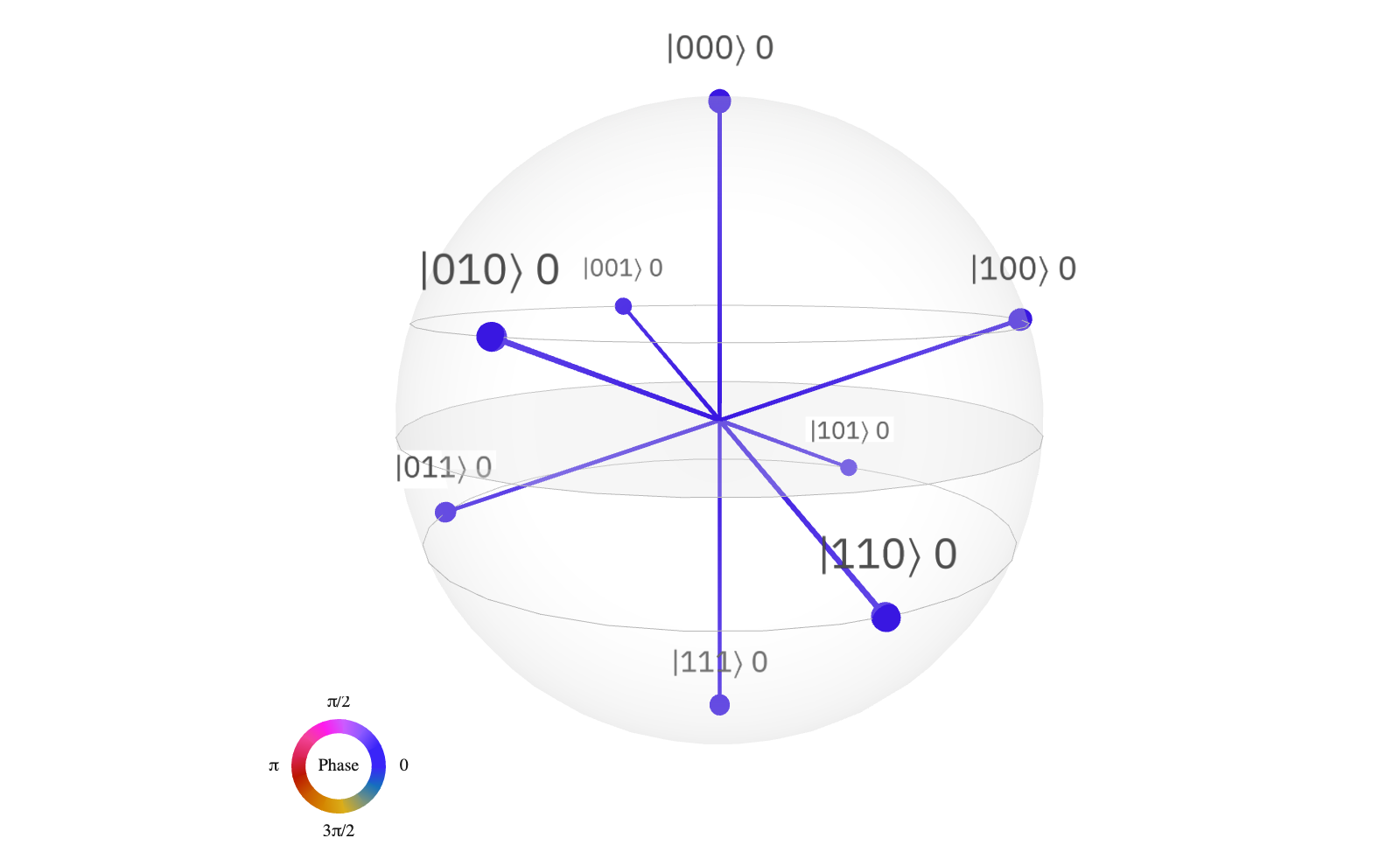}
         \caption{Visualizing the quantum state of 3 quantum bits (\qubits{}) in superposition.}
         \label{fig:summary:vis_quantum_bits}
     \end{subfigure} 
       \begin{subfigure}[t]{0.50\textwidth}
         \centering
\raisebox{0.3in}{\includegraphics[width=\textwidth]{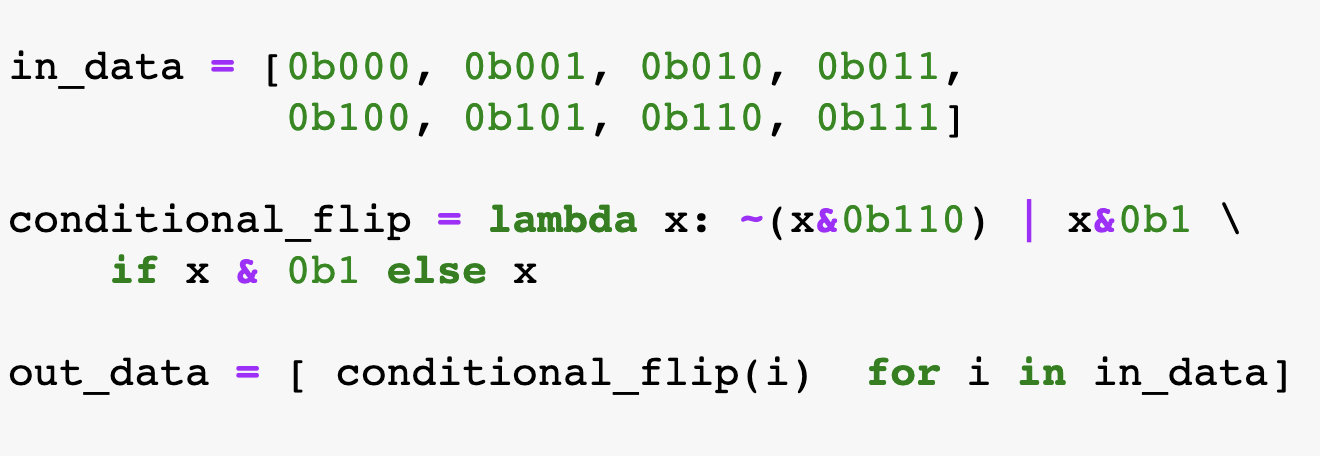}}
         \caption{Classical (python) code to flip bits 1 and 2 if bit 0 is set.}
         \label{fig:summary:classical_program}
     \end{subfigure}
     \hfill
    \begin{subfigure}[t]{0.45\textwidth}
         \centering
 \includegraphics[width=\textwidth]{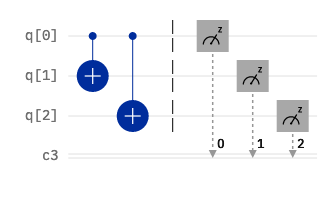}
         \caption{Quantum code (circuit) to flip the state of qubits 1 and 2 if the state of qubit 0 is $\ket{1}$}
         \label{fig:summary:quantum_circuit}
     \end{subfigure} 
     \vspace{8pt}
     \caption{Comparison of classical  quantum state visualization and programs.}
     \label{fig:classical_vs_quantum}
\end{figure*}

\section{Quantum Computing Background}
\label{sec:qc_background}


\vspace{4pt} 


The language we use to describe concepts in \qx{}, terms like \emph{qubit}, \emph{quantum state}, \emph{superposition}, and \emph{entanglement} may be unfamiliar to many.
The information contained in classical bits and quantum bits (\qubits{}) is completely different  thereby necessitating completely different visual data presentation metaphors as illustrated in Fig.~\ref{fig:summary:vis_classical_bits} and Fig.~\ref{fig:summary:vis_quantum_bits}.
The methods for programming these systems bear little resemblance to general purpose languages like C++ and Python as shown in Fig.~\ref{fig:summary:classical_program} and Fig.~\ref{fig:summary:quantum_circuit}.
Unlike the practices we saw during the serial-vector-massively parallel (MPP) evolution, there is no straightforward path for taking a serial classical code and transforming it  so that it will run on quantum platform.
%
For additional background, Sidebar \S\ref{sec:sb:getting_started} provides a set of resources that are useful in learning more about \qx{}.
%

%
%


In \qx{}, information is represented in the \emph{quantum state} of \emph{quantum bits}, or \qubits{}.
Unlike in the classical world where the state of a \emph{bit} is either a value of 0 or 1, the state of a \qubit{} is more complex owing to the nature of the  physics of quantum mechanics upon which it is constructed.
Intuitively, the \emph{quantum state} of a \qubit{} may be thought of as a combination of potential outcomes  rather than a discrete binary value. 

In the quantum world, we represent the \emph{quantum state} of a \qubit{} using \emph{Dirac} notation, which is the standard way of describing states in quantum mechanics.
Two possible states of a \qubit{} are $\ket{0}$ and $\ket{1}$, which are known as \emph{computational basis states} that 
correspond to the classical bit states of 0 and 1. 
Unlike classical bits, a \qubit{} may be in a state other than $\ket{0}$ and $\ket{1}$. 
It is possible to form a linear combination of states known as \emph{superpositions}~\cite{Nielsen:Quantum:2010}:
\begin{equation}
\ket{q} = \alpha\ket{0} + \beta\ket{1}
\label{eq:qubit-state}
\end{equation}
where $\alpha$ and $\beta$ are \emph{amplitudes} and
$\lbrace  \alpha,\beta \rbrace \in \mathbb{C}^2$.
%
Born's rule states that  in a superposition of states, the modulus squared of the amplitude of a state is the probability of that state resulting after measurement. Furthermore, the sum of the squares of the amplitudes of all possible states in the superposition is equal to 1~\cite{Hidary:2019}.
Therefore, for the state $\ket{q}$ in Eq.~\ref{eq:qubit-state}, we have  ${|\alpha|}^2 + {|\beta|}^2 = 1$ and the probability of measuring state $\ket{0}$ state is ${|\alpha|}^2$ and state $\ket{1}$ is ${|\beta|}^2$.

\begin{figure}[h!]
\centering
\includegraphics[width=0.45\textwidth]{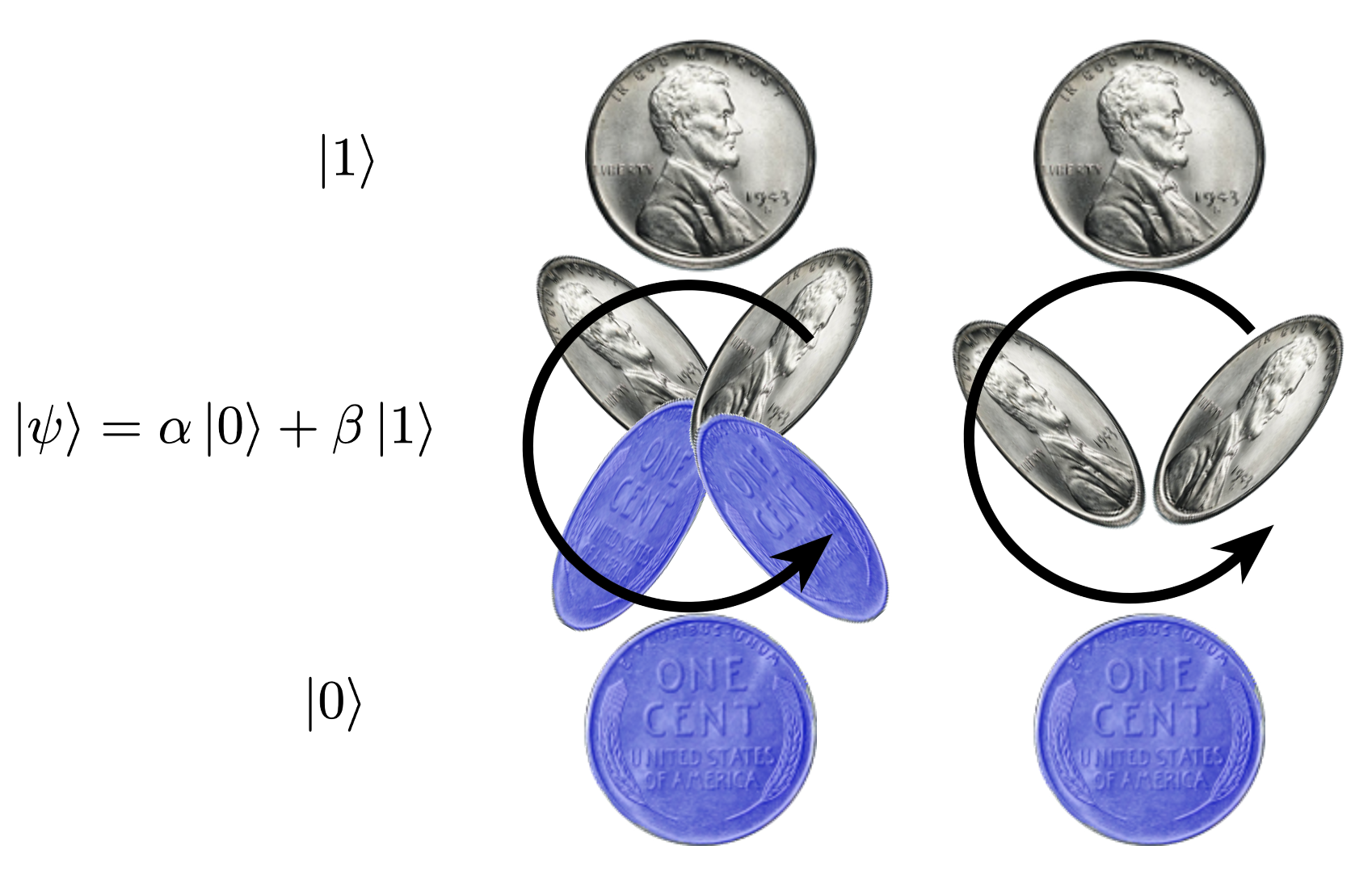}
\caption{Using a coin toss analogy,  when the coin is at rest, we have a 100\% probability of a heads state $\ket{1}$ in the top row or a tails state $\ket{0}$ in the bottom row.
When the coin is spinning  (middle row), its state $\ket{\psi}$ is a combination of probabilities $\alpha$ and $\beta$.
In the right column,  while the probability of a heads is the same for both potential coin states in the middle, their \emph{quantum phase} differs in that one head points more westerly and the other points more easterly.}
\label{fig:pennies}
\end{figure}

A useful analogy for a \emph{quantum state} is a coin toss.
While the coin is at rest and its state measured or observed, the coin is in either a heads or tails, $\ket{1}$ or $\ket{0}$ respectively as shown in the top and bottom rows of Fig.~\ref{fig:pennies}.
On the other hand, while the coin is spinning in the air, the coin is in a state of \emph{superposition} where its state is a combination of the heads and tails states as shown in the middle row of Fig.~\ref{fig:pennies}.
%
In the quantum world, once the state of the \qubit{} is observed or measured, the value we observe is either 0 or 1. 
Once we measure or observe a \qubit{} its quantum state collapses and the $\alpha$ and $\beta$ are no longer accessible. 
%

\begin{figure*}[th!]
   \centering
     \begin{subfigure}[b]{0.67\textwidth}
         \centering
         \includegraphics[width=0.8\textwidth]{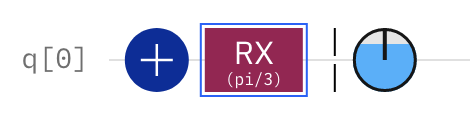}
         \includegraphics[width=0.25\textwidth]{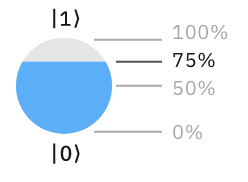}
         \caption{A quantum circuit that transforms a \qubit{} from an initial state of $\ket{0}$ first into $\ket{1}$ with a NOT gate (blue circle with a plus) then into a state of superposition by rotating the quantum state about the X axis with an RX gate by $\theta = \pi/3$. The phase disk on the right of the circuit shows the probability of a $\ket{1}$ state, and the legend below shows the probability in terms of percentage.
         }
         \label{fig:phase_disk_circuit}
     \end{subfigure}
     \hfill
%
%
     \begin{subfigure}[b]{0.30\textwidth}
     \centering
     \includegraphics[width=\textwidth]{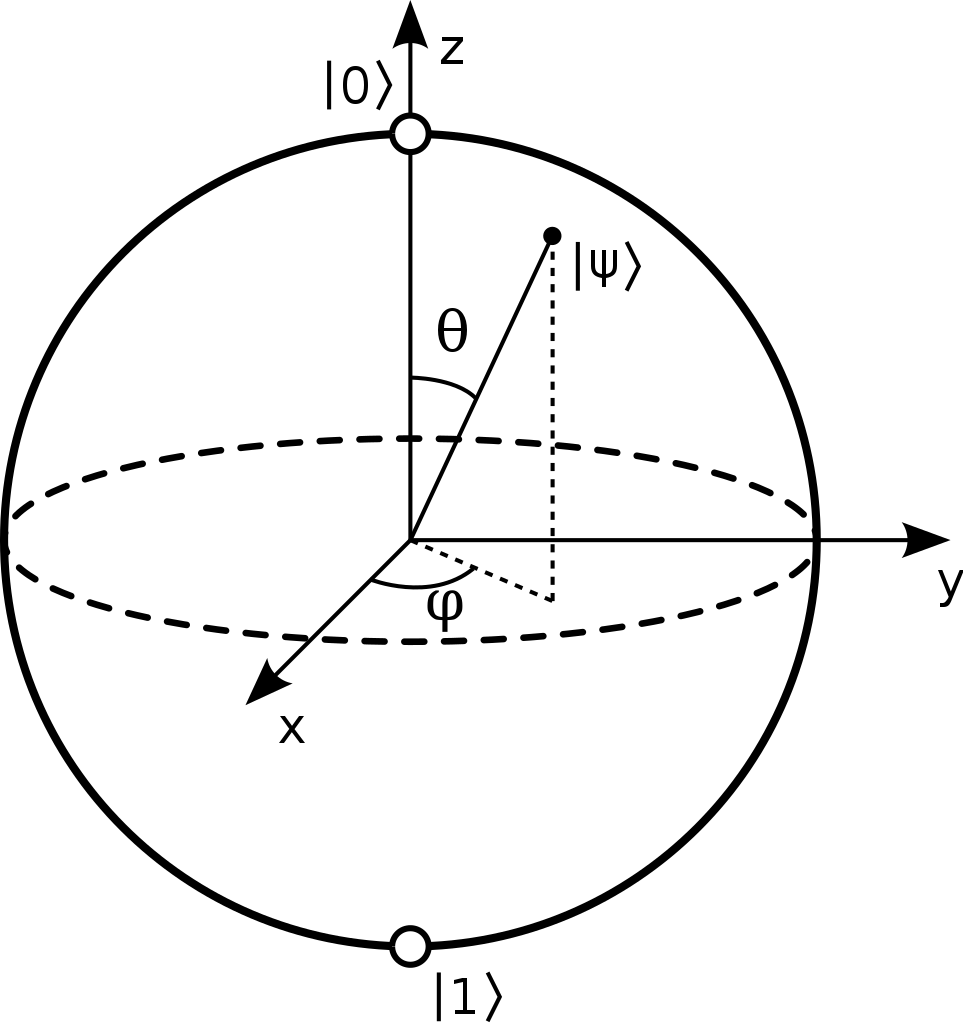}
     \caption{The \blochsphere{} maps the 3-dimensional quantum state of a single \qubit{} to a point on the unit sphere (Image source: Wikimedia Commons).}
     \label{fig:bloch_sphere}
     \end{subfigure}
\caption{ Two different visualization methods for seeing the quantum state of single qubit.  }
\label{fig:qubit_state_simple}
\end{figure*}

In addition to the idea of a probability of a heads or tails state, there is also the notion of \emph{quantum phase}. 
Consider the right column, middle row of Fig.~\ref{fig:pennies} where there are two possible configurations of a coin in a being-tossed state.
Both of these configurations appear to have the same probability of being a heads but they differ in \emph{phase} since one head points in a more westerly direction and the other points in a more easterly direction. 
While there are two different types of quantum phase, a \emph{global phase} and a \emph{relative phase}~\cite{Hidary:2019}, for the discussion here, the most important idea is that it is possible to have two (or more) quantum states that have identical probability amplitudes for a heads and tails state, but like the right side of Fig.~\ref{fig:pennies}, are completely different states.
Quantum programs provide the means to manipulate the quantum state of single or multiple entangled \qubits{} in ways that change either the \emph{probability amplitudes} of observing a $\ket{0}$ or $\ket{1}$ or the phase, or both. See Sidebar~\ref{sec:sidebar_euler_angles} for more information.


%



Contemporary quantum platforms are not quite yet as usable as classical platforms due to the fact the underlying hardware is still maturing.
Unlike classical bits comprised of flip-flops that retain their state so long as voltage is present~\cite{PattersonHennessy:2021}, \qubits{} are able to reliably retain state for only a short amount of time, typically on the order of hundreds of microseconds\cite{IBM:Quantum:Cloud:2023}.
Present day systems are referred to as \emph{Noisy Intermediate Scale Quantum} (NISQ) platforms because the quantum state in \qubits{} and gates between them is subject to interference from external sources like heat and energetic particles in the environment~\cite{Preskill2018quantumcomputingin}.
As a result, we typically will run a quantum program thousands of times to obtain a distribution of outcomes from a given computation.


Programming and the design of \qx{} algorithms are fundamentally different than the classical world. 
In the classical world, we read in data from disk then perform operations on it.
In the \qx{} world, most programs create an initial set of quantum states through a process of \emph{superposition} so that a \qubit{} is in a state that looks like Eq.~\ref{eq:qubit-state}, and then the quantum program performs operations on the quantum states, and then finally measures the \qubits{} to obtain the final answers.
The operations on quantum states must conform to a specific set of rules and constraints that may be thought of as applying a series of rotation matrices to the quantum state as further explained in Sidebar~\S\ref{sec:sidebar_euler_angles}.
Transferring classical data into quantum systems and then back out again is an area of ongoing research and engineering. 
There are several useful surveys of contemporary environments and approaches for creating and running quantum programs (c.f.~\cite{Hidary:2019}).

\section{How Visualization Helps \qx{}}
\label{sec:vis_for_quantum}

Visualization plays a central role in facilitating deeper understanding of \qx{}.
In the subsections that follow, our examples focus on visualization of the quantum state for single and multiple \qubit{} systems using methods from contemporary \qx{} programming environments.
These examples are not an exhaustive set of such visualization techniques.
Additionally, visualization has played a role in helping to understand performance data associated with quantum programs.
Such data include measures of runtime performance, reliability as a function of program size, and more.
A survey of these types of visualizations is beyond the scope of this article.



%
%

\subsection{Visualizing the Quantum State of Individual \Qubits{} }
\label{sec:vis_one_qubit}

In the classical world, since the state of a \emph{bit} is either 0 or 1, visualizing the state of a classical bit is straightforward using any number of scalar information visualization methods. 
Of historical importance is the role played by visual presentation of the state of groups of classical bits in many contexts, including presentation of type via bitmapped fonts,  glyphs (see Fig.~\ref{fig:summary:vis_classical_bits}), and animated characters in video games.
The visualization of groups of classical bits is an integral part of the modern computational landscape on virtually all display devices.

\begin{figure*}[th!]
\centering
     \begin{subfigure}[t]{0.27\textwidth}
         \centering
         \raisebox{0.25in} { \includegraphics[width=\textwidth]{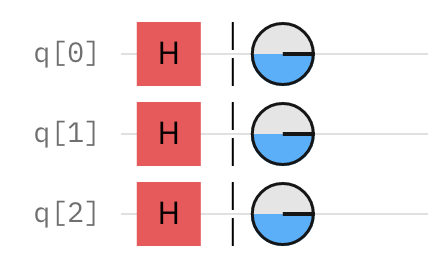} }
         \caption{A quantum circuit where each \qubit{} is put into a state of superposition. The phase disk shows equal probabilities of each state in each \qubit{}.}
         \label{fig:3q_H_gates}
     \end{subfigure}
     \hfill
     \begin{subfigure}[t]{0.39\textwidth}
         \centering
        \raisebox{0.1in} { \includegraphics[width=\textwidth]{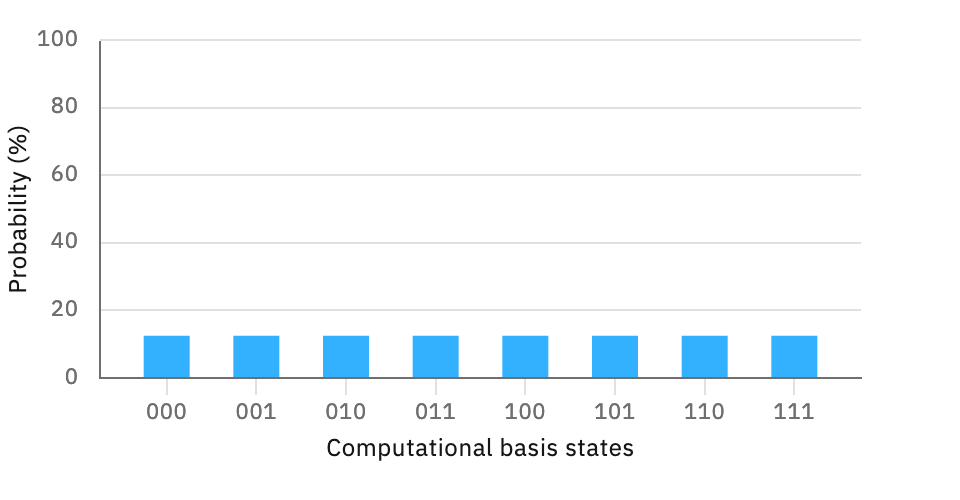} }
         \caption{Bar chart showing the (equal) probabilities of each of the quantum  states.}
         \label{fig:3q_H_bars}
     \end{subfigure}
     \hfill
    \begin{subfigure}[t]{0.30\textwidth}
         \centering
         \includegraphics[width=\textwidth]{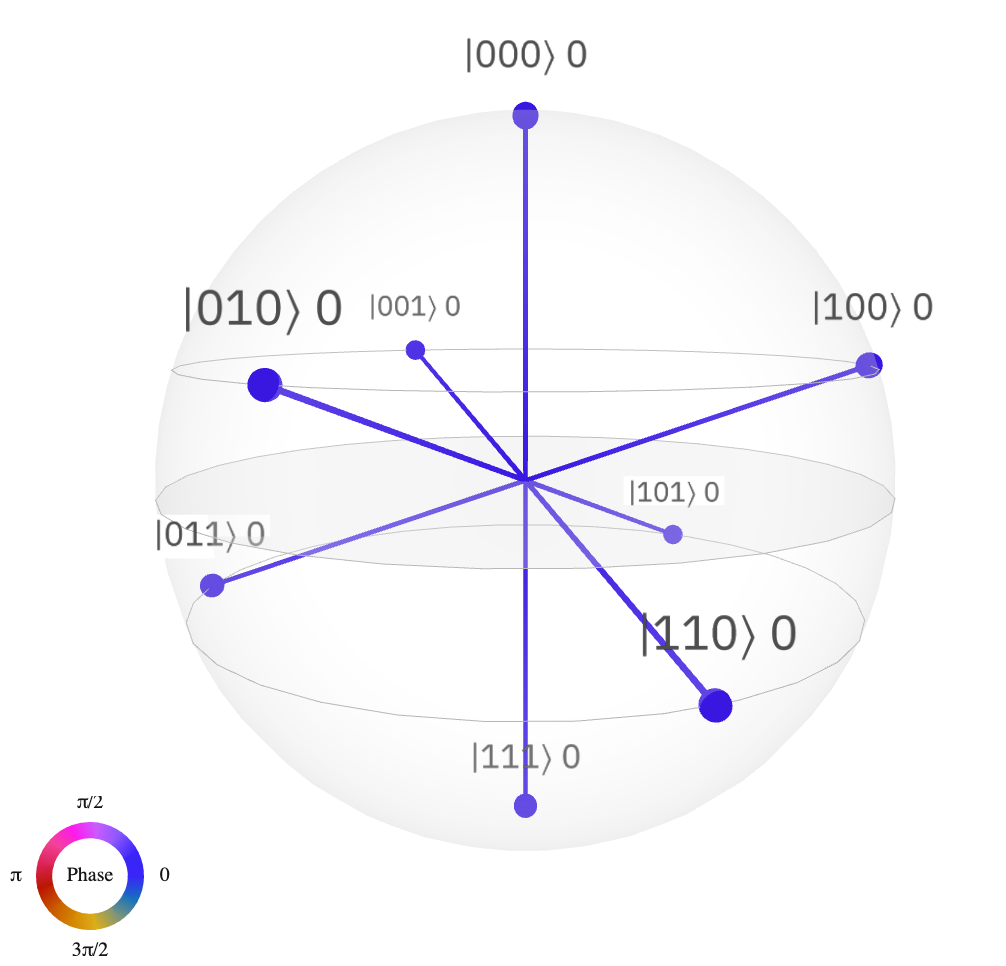}
         \caption{Q-sphere visualization showing the probabilities and phase of each of quantum states.}
         \label{fig:3q_H_qsphere}
     \end{subfigure}
\caption{An $n=3$ \qubit{} circuit (left) where the quantum state of each \qubit{} is placed in a state of superposition resulting in an overall quantum state of size $2^n$. The probabilities of each of the $2^n$ states is shown via a bar chart (middle) and in a Q-sphere format (right).} 
\label{fig:3q_h}
\end{figure*}


Visualizing the quantum state of a \qubit{} requires a different approach due to the nature of the information in the quantum state.
One type of information is the probability of obtaining a $\ket{0}$ or $\ket{1}$ state upon measurement.
%
As an example, consider the simple quantum circuit in Fig.~\ref{fig:phase_disk_circuit} where \qubit{} zero begins in state $\ket{0}$.
Then, the state is inverted with a NOT gate (the "$+$" symbol) then transformed by the RX gate, which applies a rotation transformation about the X axis here in the amount of $\theta = \pi/3$.
The result is that the quantum state has non-zero probabilities for states $\ket{0}$ and $\ket{1}$. 
At the rightmost part of the circuit is a small circle-shaped glyph that is partially filled with a shade of blue to indicate the probability of the \qubit{} being in state $\ket{1}$.
This particular glyph is known as a \emph{phase disk} and is part of the IBM Quantum Composer suite of visualization tools~\cite{IBM_Quantum_Composer:Visualizations:2023}. 

\begin{figure*}[th!]
\centering
     \begin{subfigure}[t]{0.34\textwidth}
         \centering
         \raisebox{0.25in} { \includegraphics[width=\textwidth]{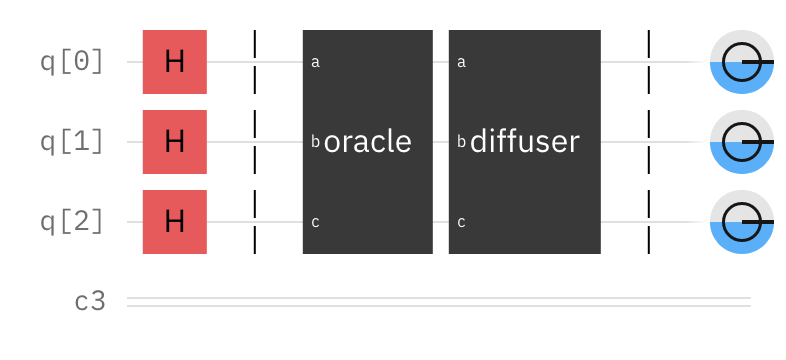} }
         \caption{The two-stage circuit consists of unitary transformations that boost the probability of the desired solution state $\ket{111}$. }
\label{fig:3q_grovers_oracle_diffuser_circuit}
     \end{subfigure}
     \hfill
     \begin{subfigure}[t]{0.34\textwidth}
         \centering
         \includegraphics[width=\textwidth]{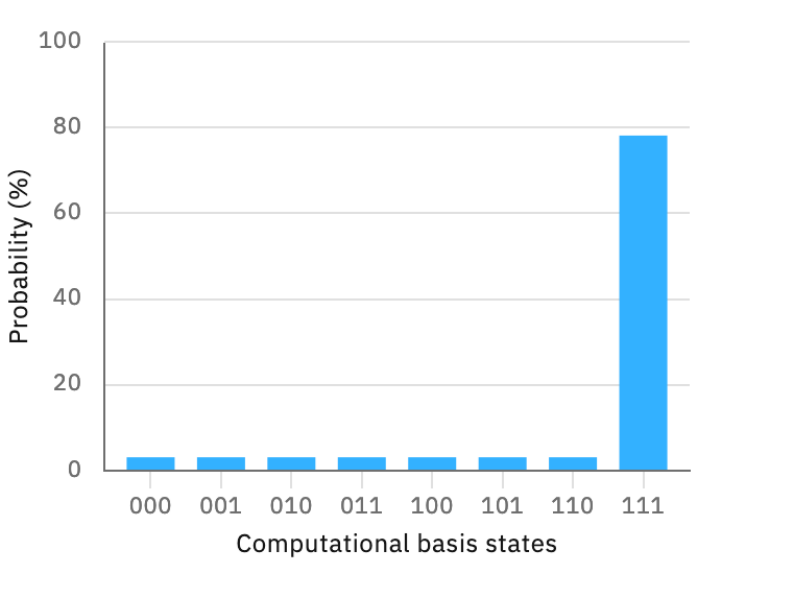}
         \caption{The histogram shows the probability distribution of all potential quantum states with state $\ket{111}$ having the highest probability.}
         \label{fig:3q_grovers_oracle_diffuser_histogram}
     \end{subfigure}
     \hfill
    \begin{subfigure}[t]{0.28\textwidth}
         \centering
         \includegraphics[width=\textwidth]{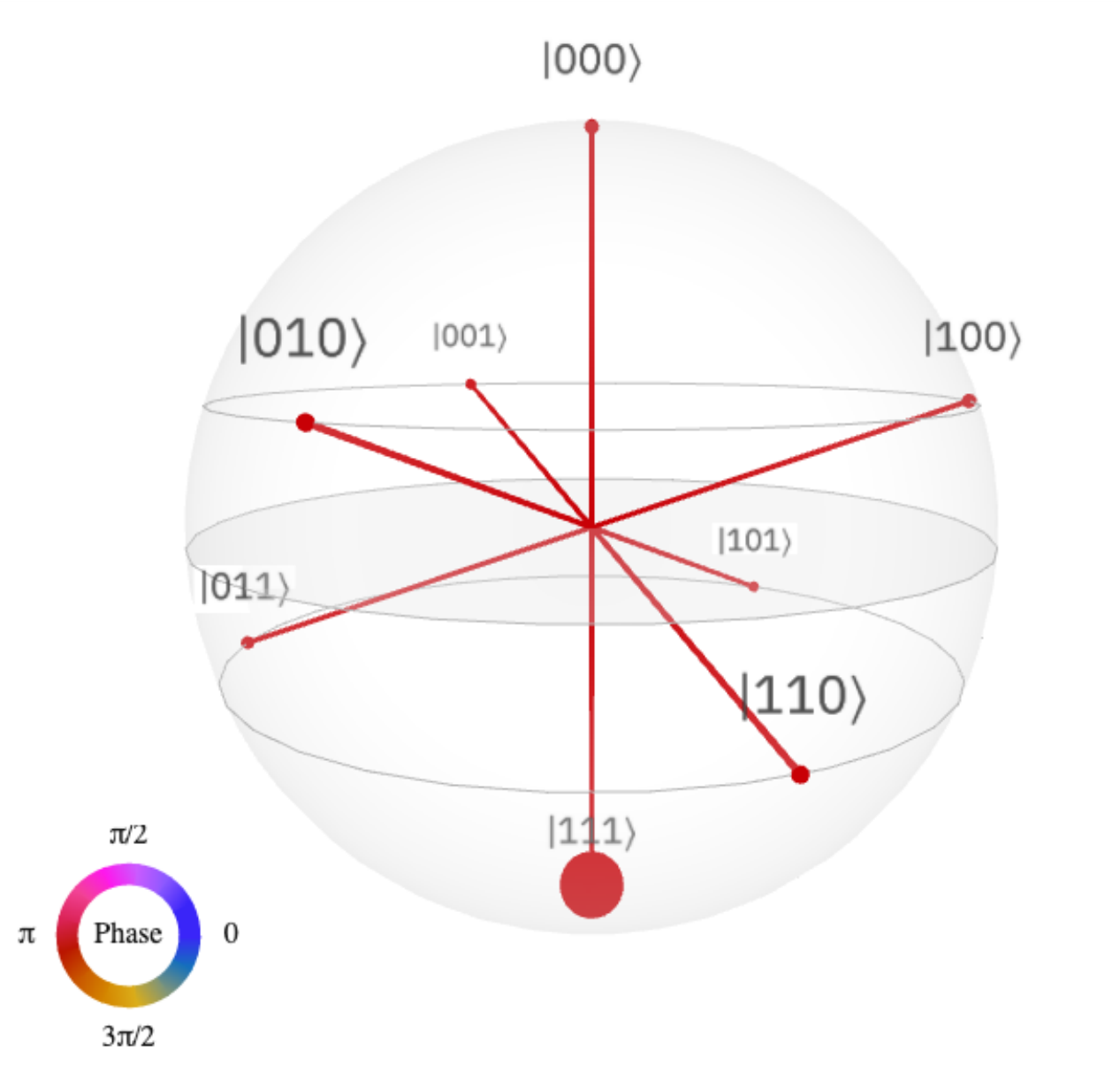}
         \caption{The Q-sphere visualization shows that all quantum states have the same phase, but state $\ket{111}$ has a higher probability than all the others.}
         \label{fig:3q_grovers_oracle_diffuser_qsphere}
     \end{subfigure}
\caption{Quantum circuit and visualization of state for Grover's Search algorithm.} 
\label{fig:3q_grovers_oracle_diffuser}
\end{figure*}

The other type of information in the quantum state is \emph{quantum phase}.
To obtain the value of phase, we can rewrite Eq.~\ref{eq:qubit-state} as  
\begin{equation}
      \ket{\psi} = cos{\theta \over 2}\ket{0} + e^{i\psi}sin{\theta \over 2}\ket{1} 
\end{equation}
Due to the fact that ${|\alpha|}^2 + {|\beta|}^2 = 1$, the real numbers $\theta$ and $\psi$ define a point on the surface of a unit-three dimensional sphere.
This representation is often visualized using a glyph known as a Bloch Sphere~\cite{Nielsen:Quantum:2010,Bloch:1946} and is shown in Fig.~\ref{fig:bloch_sphere}.
In the Bloch Sphere, the North pole corresponds to state $\ket{0}$ and the South pole to $\ket{1}$.
The $\theta$ angle represents the relationship of the probability amplitudes of $\alpha, \beta$, while the angle $\psi$ is the phase of the quantum state. 
The phase disk glyph in Fig.~\ref{fig:phase_disk_circuit}  has a black line anchored at the center of the disk.
Its orientation shows the phase of the quantum state like a compass.
\subsection{Visualizing the Quantum State of Multiple \Qubits{} }
\label{sec:vis_multiple_qubit}





If we expand our earlier coin-toss scenario from a single coin to $n$ simultaneous coins, the size of the state space describing all possible permutations of heads and tails combinations 
grows exponentially. 
This idea applies to \qubits{} as well:
if there is  one coin, or \qubit{}, there are $2^1$ possible basis states in this system $ \lbrace \ket{0}, \ket{1} \rbrace $; if there are two coins, or \qubits{}, there are $2^2$ possible basis states in this system $\lbrace \ket{00}, \ket{01}, \ket{10}, \ket{11} \rbrace$; if there are $N$ \qubits{}, then there are $2^N$ possible basis states.

The exponential growth in the size of the quantum state space is where \qx{} shows promise in being able to tackle problems of a scale not possible on classical computers.
For example, a 300 \qubit{} system can theoretically represent a quantum state of size $2^{300}$, which is more than the estimated number of particles in the visible universe and is far larger than any classical system.


For visualizing the quantum state of multiple \qubits{}, consider the example shown in Fig.~\ref{fig:3q_h}.
Three \qubits{} in a quantum circuit are placed into states of superposition using a Hadamard gate, which has the effect of creating a state of superposition where there is an equal probabilities of states meaning there is an equal likelihood of measuring either a  $\ket{0}$ and $\ket{1}$.
This gate is one of the foundational building blocks of many quantum programs~\cite{Hidary:2019}.
For this $n=3$ \qubit{} system, the size of the quantum state is $2^3=8$ and the bar chart Fig.~\ref{fig:3q_H_bars} shows the probability of each of the $2^3$ basis states.
In this case, all the bars are the same height indicating that each of these basis states $\lbrace \ket{000}, \ket{001}, \dots, \ket{111} \rbrace$ have equal probabilities.

Fig.~\ref{fig:3q_H_qsphere} shows what is known as a \emph{\qs{}} presentation of quantum state~\cite{IBM_Quantum_Composer:Visualizations:2023}. Unlike the Bloch Sphere, which shows the state of a single \qubit{}, the \qs{} shows the global quantum state by associating each of the $2^n$ 
basis states from $n$ \qubits{} with a point, or node, on the surface of the sphere. 
The radius of each node on the sphere's surface is proportional to the probability of that basis state. 
In Fig.~\ref{fig:3q_H_qsphere}, since each basis state has equal probability, all the nodes are the same size, though the perspective foreshortening results in the nodes appearing to be of varying sizes.
The color of each node indicates quantum phase of that particular basis state.
The circular color legend on the lower left provides the visual association between color and phase.  

In terms of node positioning on the \qs{} in Fig.~\ref{fig:3q_H_qsphere}, nodes are laid out so that the $\ket{000}$ and $\ket{111}$ states are at the north and south pole, respectively. 
Going from the north pole and heading southward, each successive latitude has basis states with a greater number of $\ket{1}$ states and the latitude of a given basis is determined by its Hamming distance from the $\ket{000}$ state. 
In this example, there are three states with a Hamming distance of 1 from $\ket{000}$, namely $\lbrace \ket{001}, \ket{010}, \ket{100} \rbrace$, and three states with a Hamming distance of 2 from $\ket{000}$, namely $\lbrace \ket{011}, \ket{101}, \ket{110} \rbrace$.

\section{How \qx{} Helps Visualization}
\label{sec:quantum_for_vis}


In 2001, A. Glassner presented a 3-part column in IEEE CGA~\cite{Glassner:2001} that provided an excellent introduction and overview to \qx{} and also began exploration of how \qx{} might be used in graphics and visualization.
Langzagorta et al.,~\cite{Lanzagorta:2003} suggested two broad approaches when considering ways to leverage \qx{} for graphics.
The first is to identify portions of a classical algorithm that are similar to known methods in \qx{}.
%
%
The examples we present are works that pursue this approach.
The second and much more difficult approach is to find structure in the classical algorithm that can fully exploit the quantum nature of quantum platforms, namely the ability to perform an operation in parallel across all dimensions of a quantum state. 
%
%

We begin this section with a brief description of a commonly used approach in the design of \qx{} methods where a quantum operation is applied in parallel to all basis states of the quantum state.
Then, we present some examples of visualization and related methods that take advantage of \qx{}.


\subsection{\qx{} Algorithms and the Quantum State}

\qx{}s power arises from the exponential growth in the size of the quantum state with increasing numbers of \qubits{} combined with the ability to apply an operation simultaneously to all quantum states leading to an exponential speedup for certain types of computations. 
Many quantum programs begin by placing individual \qubits{} into states of superposition with operations like the Hadamard gate~\cite{LaPierre:2021}  as shown in 
Fig.~\ref{fig:3q_H_gates}.
Next, the quantum program 
operates \emph{in parallel} on all quantum basis states using a function  $U_f$ that is a \emph{unitary} implementation of some classical function $f$.

Unitary operators have special properties; they may be thought of as rotation matrices that manipulate the quantum state (see Sidebar~\ref{sec:sidebar_euler_angles}).
The input to $U_f$ is $\ket{x}$, and its output is $U_f\ket{x}$, which is $U_f$ evaluated $f$ on all possible values of the input $x$.
The quantum circuit for such an operation is shown in Fig.~\ref{fig:3q_grovers_oracle_diffuser_circuit} where two groups of $U_f$ unitary operators, shown as \emph{oracle} and \emph{diffuser}, operate in parallel on all quantum states.
The impact of this type of computational model is profound and captures the essence of the power of \qx{}: we can evaluate some $f(x)$ over all $2^n$ quantum states of $n$ entangled \qubits{} in a single operation.
Not shown in Fig.~\ref{fig:3q_grovers_oracle_diffuser_circuit} is how the quantum program will create a state of entanglement between one or more \qubits{} through the use of unitary operations like the Controlled NOT gate (CNOT)~\cite{LaPierre:2021} which conditionally inverts the state of one \qubit{} based upon the state of another.

One of the foundational quantum algorithms is Grover's Search (c.f.~\cite{LaPierre:2021}) for finding a item or items in an unordered list of length $N$.
A classical implementation of an unstructured search requires on average $O(N/2)$ steps since on average about half of the items in the set would need to be examined.
The circuit for Grover's Search is shown in Fig.~\ref{fig:3q_grovers_oracle_diffuser_circuit} for a 3-\qubit{} system.
The intention of the two unitary transforms \emph{oracle} and \emph{diffuser} is  to boost the probability of solution states and reduce the probability of non-solution states.
Those probabilities are shown as a histogram in Fig.~\ref{fig:3q_grovers_oracle_diffuser_histogram} and in Q-sphere format in Fig.~\ref{fig:3q_grovers_oracle_diffuser_qsphere}.

The process of finding the set of unitary operations that manipulate the $2^n$ basis states of an $n-$\qubit{} system is a non-trivial process.
For this reason, nearly all work that has attempted make use of \qx{} for graphics and visualization has focused on mapping a portion of a classical algorithm to make use of Grover's Search.



\subsection{\qx{} and Visualization }

Glassner's 2001 column~\cite{Glassner:2001} provided an overview of how one might create a quantum radiosity rendering algorithm.
The central idea is to encode subsets of the scene into \qubits{} storing information like  the amount of light being emitted, the amount of incident light, the amount of reflected light and so forth.
Then, once the state of the scene is encoded into \qubits{}, to use a variant of Grover's Search to find the configuration with minimum energy, which corresponds to a solution to the radiosity equation.
This notional overview was impractical since it was estimated to require hundreds of millions of bits assuming a 5,000 surfaces and 32-bit resolution for encoding information.
%

%
%
%
%


%
Lanzagorta et al., 2003~\cite{Lanzagorta:2003} suggest using Grover's to accelerate searches of objects in a graphics database, a z-buffer algorithm to determine which object is closest to a viewer, and a ray-object intersection methodology.
More recent work in this space~\cite{TowardsQuantumRayTracing:2022} builds on this idea and includes an implementation that runs on a 20-\qubit{} system.
The key idea is to encode the scene and ray parameters into the quantum state then leverage Grover's to find ray-object intersections.
%
This idea is illustrated in Fig.~\ref{fig:summary:classical_vs_quantum_ray_tracing}.
While Grover's offers computational advantages for unstructured search, encoding classical data on into quantum state is a non-trivial process and must take into account the limitations of the size of the circuit that may be reliably executed on a a given quantum platform. 


%
A quantum version of the RANSAC algorithm~\cite{QuantumRandomSampleConsensus:2009} uses a Grover's variant to reduce computational complexity from $O(N)$ to $O(\sqrt{N})$. RANSAC fits models to data and uses many searches to find data that satisfy a set of criteria, e.g., maximums. 
This type of algorithm has uses in geometric modeling and vision problems, including camera pose estimation, structure from motion, and shape detection.
%


There are fewer examples of work in visualization that follow the more difficult approach where some or all of a classical algorithm is refactored for the quantum environment.
One example that takes steps in that direction is in the area of pixel supersampling~\cite{Johnston:2016} using multiple quantum processing stages to produce what appear to be better visual results than an approach based on classical Monte Carlo sampling.
These different results are a direct result of how a quantum sampling approach results in a significantly different probability distribution compared to a classical approach (Fig.~\ref{fig:summary:classical_vs_quantum_walk}).

\section{The Road Ahead}

Despite the examples here that suggest \qx{} may be able to tackle larger problems and produce better results than are possible on classical platforms, there are a number of significant challenges. 

From a visualization-helps-quantum perspective, the examples we have shown in this article 
suggest that most visualization methods are useful for only a small number of \qubits{}. The existing set of methods and approaches do not seem promising for use on configurations with dozens or hundreds of \qubits{}, particularly considering that the size of the quantum state scales exponentially with the number of \qubits{}.
Additionally, there is much less work in the area of visualizing the behavior of large, entangled systems of \qubits{}. 

For the quantum approaches we have described, there is a common theme: initialize some \qubits{}, place them into superposition, entangle them, then apply a unitary operator $U_f$ to all quantum states at once. What is missing from this picture is somehow moving classical data into the quantum machine and then obtaining answers that are then returned to the classical world. 
A significant issue when designing and implementing quantum algorithms that process classical data is the \emph{data encoding problem}, which refers to how data is encoded in the quantum state of a \qubit{} register.
Some recent work explores approaches for encoding gray-level classical pixel data on quantum platforms~\cite{Amankwah:NatureScientificReports:2022} through a process using gate-level operations to set up the quantum state to known values.
Current quantum platforms have a limited number of \qubits{} and have limited \emph{quantum volume}, a term that refers to the number of program steps (gates) that can reliably run across \qubits{}.
Data encoding methods are not "for free": they increase the volume of the quantum circuit. 
Since current quantum platforms are of $O(10)-O(100)$ \qubits{} (and quantum annealing systems like D-Wave's approaching $O(1000)$ of \qubits{}), it is unclear how we will go about encoding "large data sets" for the purpose of analysis and visualization on quantum platforms. 
We will need to rethink our "large data visualization" approaches for quantum platforms. 

In his 2021 Turing Award presentation, J. Dongarra suggested that \qx{} platforms will likely be accelerators on future systems, at least in the near- to medium-term.
These quantum-based accelerators will co-exist with classical computing infrastructure resulting in a hybrid system.
It may be the case that the computational workload will be partitioned across classical and quantum machines much like we see a partitioning of workloads across CPU and GPU. 
Such \emph{hybrid architectures} will likely figure prominently in successful approaches for quantum-leveraging visualization and graphics methods in the near- and medium-term.

\section{Conclusion}

The field of \qx{} is "up and coming", and has potential to tackle problems that far exceed what is possible on classical platforms.
It is a disruptive technological change that will require significant rethinking of how we formulate and implement algorithms.
The focus of this article has been to highlight some synergies between visualization and \qx{} both in terms of how visualization helps with understanding \qx{} and how visualization might benefit from use of \qx{} formulations of problems.

\section{Sidebar: Getting Started with \qx{}}
\label{sec:sb:getting_started}

\vspace{4pt}  

If you are new to the world \qx{} and would like to learn more, there are a number of useful resources providing additional breadth and depth beyond the scope of the article.
A good start is Andrew Glassner's 3-part column in IEEE CGA from 2001, which provides a well rounded overview of \qx{} along with some speculation of how it might be used in graphics~\cite{Glassner:2001}.

Several excellent learning resources are developed and distributed online and at no cost by quantum platform companies in the interest of increasing awareness and expertise in \qx{} topics.
Some examples include  IBM Quantum and their Qiskit platform~\cite{Qiskit:Textbook:2023}, Xanadu.ai and their PennyLane platform~\cite{Xanadu:Cookbook:2023}, Google's Quantum AI and their Cirq platform~\cite{Google:QuantumQI:Cirq:2023}.
This set is not exhaustive nor complete, there are numerous additional resources accessible via the web.

\emph{Online tutorials.} 
Qiskit resources~\cite{Qiskit:Tutorials:2023} include a "\qx{} in a nutshell" as a good starting point along with a link to a YouTube channel with many useful videos that present quantum topics in an accessible way. 

\emph{Coding environments.}
Some good resources for learning how to write quantum programs are online at the Qiskit~\cite{Qiskit:Tutorials:2023}, Xanadu~\cite{Xanadu:Cookbook:2023}, and Cirq~\cite{Google:QuantumQI:Cirq:2023} sites.
These provide information about how to create quantum circuits, create states of superposition and entanglement, measure and visualize quantum state, and execute quantum programs on simulators and actual quantum hardware.
The examples cited here are accessible as freely accessible Python modules. Code development consists of problem setup in Python using the specific capabilities of a particular \qx{} coding platform.

\emph{Cloud-based platforms.}
With your first \qx{} programs you will likely be running them via Python on a local \emph{quantum simulator} that is included as part of the \qx{} coding environment. 
When you are ready to try running on actual hardware, there are several different cloud-based providers of quantum platforms. Some of these systems are accessible at no cost while others use a fee-based access model. 
IBM Quantum~\cite{IBM:Quantum:Cloud:2023} provides access to about two dozen systems ranging in size from 5 to 433 \qubits{} and is well integrated with Qiskit's runtime system.
Microsoft's Azure Quantum service~\cite{Microsoft:Azure:Quantum:2023} and Amazon's Braket~\cite{Amazon:Braket:2023} are both gateways to quantum platforms from IonQ, Quanntinuum, Rigetti, and others ranging in size from 10-80 \qubits{} and supports running codes written in Qiskit, Cirq, PennyLane, and others.

\section{Sidebar: Quantum Operations as Rotation Matrices}
\label{sec:sidebar_euler_angles}


Quantum operations on $n$ qubits $\ket{\psi}$ are described in the langauge of mathematics as $2^n \times 2^n$ \emph{unitary matrices}. In particular, the quantum state resulting after applying a quantum operation $U$ to a quantum state $\ket{\psi}$ is given by matrix multiplication
\begin{equation}
    \ket{\psi'} = U \ket{\psi} \,.
\end{equation}
Properties of quantum computation are consequences of unitary computation. For example, quantum states are closed under quantum operations since a unitary matrix $U$ preserves the norm of an input vector so that $\ket{\psi'}$ also has unit norm. As another example, a unitary matrix has the property that $UU^\dagger = U^\dagger U = I$ where $U^\dagger$ is the conjugate transpose of $U$. This means that every quantum operation is \emph{invertible} or \emph{reversible} since the definition implies that $U^{-1} = U^\dagger$. Notably, classical computation does not have this property. 

In practice, one cannot write a quantum program by specifying all the coefficients in the unitary matrix $U$ since it is of size $2^n \times 2^n$, which is exponential in the number of qubits $n$. Instead, quantum hardware devices supply primitive \emph{gates} from a finite set of gates which can be sequenced together to form a \emph{quantum circuit}. Each quantum circuit corresponds to a unitary matrix. Thus a quantum circuit enables us to express a quantum computation in a compact form that is physically implementable without explicitly specifying all $2^n \times 2^n$ coefficients. As with classical computing, there are universality results in quantum computing that show that every unitary matrix can be approximated arbitrarily accurately as a quantum circuit provided that we have a universal set of quantum gates. To build some intuition for how a quantum operation (quantum circuit) works, let us highlight how it is possible to think of them as ``generalized rotations".

Consider a $1$-qubit system so that we can visualize the quantum state on a Bloch sphere. As a reminder, a $1$-qubit system can be specified as
\begin{equation}
\label{eq:qubit-state-2angles}
    \ket{\psi} = cos{\theta \over 2}\ket{0} + e^{i\phi}sin{\theta \over 2}\ket{1}
\end{equation}
on the Bloch sphere where $\theta$ is the polar angle and $\phi$ is the azimuthal angle. Since every $1$-qubit quantum state lies on the Bloch sphere, we can visually reason that a quantum operation can be expressed as a $3$-dimensional rotation in Euclidean space. Because we can use \emph{Euler angles} to represent all such $3$-dimensional rotations, we further reason that we need $3$ degrees of freedom to specify a quantum operation. At this point, the astute reader may observe an issue with our description so far that serves to highlight the quirks of quantum computation. In particular, we seem to be missing $1$ degree-of-freedom in our description since $2 \times 2$ unitary matrices have $4$ degrees of freedom whereas $3$ dimensional rotation matrices only have $3$ degrees of freedom.

The missing degree of freedom is the \emph{global phase} $\phi_g$ and can be seen by noting that $e^{i\phi_g} \ket{\psi}$ is indistinguishable from $\ket{\psi}$ by measurement. Graphically, we can see this since measurement in the computational basis gives either $\ket{0}$ or $\ket{1}$ in proportion to the distance of the vector $\ket{\psi}$ to $\ket{0}$ or $\ket{1}$ so that any rotation in the $xy$-plane of the Bloch sphere does not change the distance. The global phase is usually omitted since it is indistinguishable and the \emph{relative phase} $\phi$ is what is kept. This leads to an idea that quantum algorithms can take advantage of which is to store information in the relative phase (e.g., see quantum phase estimation which is the building block of Shor's algorithm, c.f.~\cite{Hidary:2019}).

The analogy between generalized Euler angles and unitary matrices can be made formal. On $1$ qubit systems, every unitary matrix can be written as 
\begin{equation}
    U = e^{i\phi_g} e^{iZ\alpha_1}e^{iY\alpha_2}e^{iZ\alpha_3}
\end{equation}
where $\alpha_1, \alpha_2, \alpha_3$ are generalized Euler angles, $\phi_g$ is a global phase, and $Y$ and $Z$ are Pauli matrices that correspond to Pauli gates~\cite{Nielsen:Quantum:2010}. Moreover, explicit generalized Euler angle parameterizations for general unitary matrices of all dimensions are given by Tilma et al.~\cite{tilma2004generalized}. Although the $n$-qubit case is difficult to visualize, it does shed some light on the intricacies of quantum operations, and in particular, the concept of \emph{entanglement}.

Conceptually, entanglement enables local computations on certain qubits to affect other qubits. This is unlike classical computation again, where computations on bits is local to the bits that are operated on. The phrase ``spooky action at a distance" commonly attributed to Einstein describes this counterintuitive (at least classically) consequence of entanglement. The generalized rotation view of a quantum operation helps make sense of this situation. More concretely, we might imagine that a sequence of rotations, each restricted to a single plane as with Euler angles so that at most two qubits are operated on at any given time, can still globally affect the entire quantum state, provided that the rotations are sequenced accordingly. Naturally, we are pushing the visual analogy a bit far in the context of this article to given intuition and not as a substitute for the underlying physics and mathematics.















\ifCLASSOPTIONcompsoc
  \section*{Acknowledgments}
\else
  \section*{Acknowledgment}
\fi

This research was partially supported by the U.S. Department of Energy (DOE) under Contract No.~DE-AC0205CH11231, through the Office of Advanced Scientific Computing Research (ASCR) Exploratory Research for Extreme-Scale Science Program.
The images of quantum circuits and \qubit{} state visualizations were created using the Qiskit-Terra software~\cite{Qiskit:2023} and IBM's Quantum Composer~\cite{IBM_Quantum_Composer:Visualizations:2023}.

\ifCLASSOPTIONcaptionsoff
  \newpage
\fi

\bibliographystyle{IEEEtran}
\bibliography{main}

\begin{IEEEbiography}{E. Wes Bethel}{\,}is an Associate Professor of Computer Science at \emph{San Francisco State University} and a Research Affiliate at \emph{Lawrence Berkeley National Laboratory}. Contact him at ewbethel@sfsu.edu. 
\end{IEEEbiography}

\begin{IEEEbiography}{Mercy G. Amankwah}{\,}is a Computer Science PhD student at \emph{Case Western University}. Contact her at mercy.amankwah@case.edu. 
\end{IEEEbiography}

\begin{IEEEbiography}{Jan Belewski}{\,}is a Computer Systems Engineer at the \emph{National Energy Research Scientific Computing Center} at \emph{Lawrence Berkeley National Laboratory}. Contact him at balewski@lbl.gov. 
\end{IEEEbiography}

\begin{IEEEbiography}{Roel Van Beeumen}{\,}is a Research Scientist at \emph{Lawrence Berkeley National Laboratory}. Contact him at rvanbeeumen@lbl.gov. 
\end{IEEEbiography}

\begin{IEEEbiography}{Daan Camps}{\,}is a Computer Systems Engineer at the \emph{National Energy Research Scientific Computing Center} at \emph{Lawrence Berkeley National Laboratory}. Contact him at dcamps@lbl.gov.
\end{IEEEbiography}

\begin{IEEEbiography}{Daniel Huang}{\,}is an Assistant Professor of Computer Science at \emph{San Francisco State University} and a Research Affiliate at \emph{Lawrence Berkeley National Laboratory}. Contact him at danehuang@sfsu.edu.
\end{IEEEbiography}

\begin{IEEEbiography}{Talita Perciano}{\,}is a Research Scientist at \emph{Lawrence Berkeley National Laboratory}. Contact her at tperciano@lbl.gov. 
\end{IEEEbiography}

\end{document}